\documentclass[final,5p,fleqn]{elsarticle}
\usepackage{epsfig}
\usepackage{subfigure}
\usepackage[utf8]{inputenc}
\usepackage{float}
\usepackage{graphicx}
\usepackage{amsmath}
\usepackage{verbatim}
\usepackage{xcolor}
\usepackage{caption}
\usepackage{color}
\usepackage{soul}
\usepackage{mhchem}
\usepackage{amssymb}

\def\simge{\mathrel{%
       \rlap{\raise 0.511ex \hbox{$>$}}{\lower 0.511ex \hbox{$\sim$}}}}
\def\simle{\mathrel{
       \rlap{\raise 0.511ex \hbox{$<$}}{\lower 0.511ex \hbox{$\sim$}}}}

\newcommand \beq{\begin{eqnarray}}
\newcommand \eeq{\end{eqnarray}}

\usepackage[colorlinks=true,linktocpage=true,linkcolor=blue,citecolor=blue]{hype
rref}

\usepackage{color}
\usepackage[normalem]{ulem}
\renewcommand\sout{\bgroup \color[rgb]{0.55,0.00,0.99} \ULdepth=-.5ex \ULset}

\begin{document}

\begin{frontmatter}

\title{Probing Pion Valence Quark Distribution with Beam-charge Asymmetry of Pion-induced $J/\psi$ Production }

\author[a]{Wen-Chen Chang\fnref{email1}}
\fntext[email1]{changwc@phys.sinica.edu.tw}

\author[b]{Marco Meyer-Conde\fnref{email2}}
\fntext[email2]{marco@tcu.ac.jp}

\author[c]{Jen-Chieh Peng\fnref{email3}}
\fntext[email3]{jcpeng@illinois.edu}

\author[d]{Stephane Platchkov\fnref{email4}}
\fntext[email4]{Stephane.Platchkov@cern.ch}

\author[e]{Takahiro Sawada\fnref{email5}}
\fntext[email5]{sawada@icrr.u-tokyo.ac.jp}

\address[a]{Institute of Physics, Academia Sinica, Taipei 11529,
  Taiwan}

\address[b]{Research Center for Space Science, Tokyo City University, 
Kanagawa 224-8551, Japan}

\address[c]{Department of Physics, University of Illinois at
  Urbana-Champaign, Urbana, IL 61801, USA}

\address[d]{IRFU, CEA, Universit\'{e} Paris-Saclay, 91191
  Gif-sur-Yvette, France}

\address[e]{Institute for Cosmic Ray Research, The University of
  Tokyo, Gifu 506-1205, Japan}


\begin{abstract}
We consider the beam-charge asymmetry of the $J/\psi$ production cross
sections in $\pi^-$- versus $\pi^+$-induced reactions on proton or
nuclear targets.  We show that the $J/\psi$ production cross section
difference between $\pi^-$ and $\pi^+$ beams impinging on a proton
target has a positive sign with a magnitude proportional to the
product of pion's valence quark distribution, $V_\pi$, and proton's up
and down valence quark distribution difference, $u^V - d^V$.  The
existing $J/\psi$ production data for $\pi^- + p$ and $\pi^+ + p$ at
39.5 and 200 GeV/c are consistent with the expected positive
beam-charge asymmetry. The magnitude of the asymmetry is compared with
calculations performed within two theoretical frameworks, the Color
Evaporation Model (CEM) and the Non-Relativistic QCD (NRQCD)
formalism. We also examine the beam-charge dependence for pion-induced
$J/\psi$ production cross sections measured on the neutron-rich
platinum target, and find good agreement between the data and theory
for both the negative sign and the magnitude of the beam-charge
asymmetry. The comparison between data and theoretical calculations
for both proton and platinum targets suggests that the beam-charge
asymmetry in pion-induced $J/\psi$ production is a viable method of
accessing the valence quark distribution of the pion.

\end{abstract}

\end{frontmatter}


The valence quark distributions of the proton have been extensively studied
in deep inelastic scattering (DIS) experiments. 
As the first moment of the valence quark distribution is fixed by 
the number sum rule, the experimental and theoretical interest
is mainly focused on the flavor dependence of the valence quark 
distribution~\cite{close73,farrar}.
Experimental information on $d(x)/u(x)$ at large $x$, where $x$ is 
the Bjorken-$x$, primarily comes from 
measurements of the $F^n_2/F^p_2$ neutron to proton
structure function ratio in DIS experiments.
Recent attempts to extract $F^n_2/F^p_2$ include tagged-DIS 
measurement from the BONuS experiment~\cite{CLAS12,CLAS14} and
DIS from $^3$H and $^3$He nuclei in the MARATHON experiment~\cite{Marathon}.

Compared to the proton,
the valence quark distribution of the pion, which plays an important dual
role of being the lightest meson and a Goldstone
boson, is much less known. Significant interest in the
pion substructure is reflected in the numerous theoretical predictions on
pion's valence quark distribution based on 
various approaches~\cite{Roberts_Holt,Horn}. 
The existing experimental information on its valence quark distribution
comes primarily from pion-induced Drell-Yan~\cite{DY} experiments carried out
in the 1980s~\cite{WA39,NA3DY,NA10DY,E326,E615}. 
The electromagnetic
nature of the Drell-Yan process implies a relatively small cross section,
resulting in marginal statistical accuracy for these experiments.
Moreover, the existing Drell-Yan data do not yet allow a 
separation of the valence from the sea quark distributions 
in the pion~\cite{Chang13}. 

As noted in~\cite{londergan}, the pion valence-quark distribution could be
extracted from a measurement of the difference
between the $\pi^-$ and $\pi^+$-induced Drell-Yan cross sections on 
the isoscalar deuteron ($D$) target. In leading-order formalism, 
the difference between the Drell-Yan cross 
sections, $\sigma_{DY}$, for $\pi^- + D$ and $\pi^+ + D$ reactions 
is given as
\begin{eqnarray}
\sigma_{DY}(\pi^- + D) - \sigma_{DY}(\pi^+ + D) \propto V_\pi(x_1) V_N(x_2),
\label{eq:eq1}
\end{eqnarray}
where $x_1$ and $x_2$ refer to hadron's momentum fraction carried by the
partons in the beam and target, respectively, while 
$V_\pi$ and $V_N$ are the individual valence quark
distributions in the pion and in the nucleon:
\begin{eqnarray}
V_\pi(x) & = & u^V_{\pi^+}(x)=\bar d^V_{\pi^+}(x)=d^V_{\pi^-}(x)=\bar u^V_{\pi^-}(x);
\nonumber \\
V_N(x) & = & [u^V_p(x) + d^V_p(x)]/2;
\label{eq:eq2}
\end{eqnarray}
Equation (1) shows that the 
$\pi^- + D$ and $\pi^+ + D$ Drell-Yan cross section difference is proportional
to the product of the pion and nucleon valence quark distributions. Since
the nucleon valence quark distribution is quite well known over 
a broad range of $x$ (except at very large $x$), this cross section
difference allows a direct measurement of pion's valence quark distribution.
Although the NA10 Collaboration~\cite{NA10} has measured the $\pi^- + D$
Drell-Yan cross sections, there exists no corresponding data for the
$\pi^+ + D$ reaction. While new Drell-Yan data on $\pi^+ + D$ could be
collected, for example, at the AMBER experiment~\cite{Amber} 
in the future, it is interesting to 
consider other experimental observables which are also sensitive to  
pion's valence quark distribution.

In this paper we study the possibility of probing pion's valence quark 
distribution using $J/\psi$ production with $\pi^-$ and $\pi^+$ beams.
We first show that the cross section
difference between $\pi^- + p$ and $\pi^+ + p$ $J/\psi$ production 
is proportional to the product of the pion's $V_\pi$ 
and proton's $u_V-d_V$ valence quark distributions.
We then compare the existing data on $\pi^- + p$ and $\pi^+ + p$
$J/\psi$ production with theoretical calculation.
We also examine the beam-charge dependence
for pion-induced $J/\psi$ production cross sections on the
neutron-rich platinum target.
The good agreement between data and calculation
suggests that this is a viable method to measure the valence quark
distribution of pion.  

In contrast to the electromagnetic Drell-Yan process, hadronic $J/\psi$
production involves strong interaction as the underlying mechanism.
At leading order, the two hard processes responsible for producing
a pair of $c \bar c$
quarks are the $q \bar q$ annihilation and the
$g g$ fusion. The cross section for producing a
pair of $c \bar c$ quarks is

\begin{eqnarray}
\frac {d\sigma}{dx_F d\tau} = \frac {2 \tau} {(x_F^2+4\tau^2)^{1/2}}
H_{BT}(x_1,x_2;m^2),
\label{eq:eq3}
\end{eqnarray}
where $x_1,x_2$ are the momentum fractions carried by the beam
($B$) and target ($T$) partons, and $x_F = x_1 - x_2$. The mass of
the $c \bar c$ pair is $m$, while $\tau^2 = m^2/s$ and $s$ is the
center-of-mass energy squared.
$H_{BT}$ is the convolution of the hard-process
cross sections and the parton distributions in the projectile and
target hadrons
\begin{eqnarray}
H_{BT}(x_1,x_2;m^2) = G_B(x_1)G_T(x_2)\sigma(gg \to c \bar c; m^2) \nonumber \\
+ \sum_{i=u,d,s} [{q^i_B(x_1) \bar q^i_T(x_2)+\bar q^i_B(x_1)q^i_T(x_2)}] \sigma(q \bar q \to c \bar c;m^2),
\label{eq:eq6}
\end{eqnarray}
where $G(x),q(x),$ and $\bar q(x)$ refer to the gluon, quark, and antiquark
distribution functions, respectively. The expressions for the
cross sections of the QCD subprocesses $\sigma (gg \to c \bar c)$
and $\sigma (q \bar q \to c \bar c)$ can be found, for example,
in Ref.~\cite{combridge}.

In the color-evaporation model (CEM)~\cite{cem}, the $J/\psi$
production cross section is obtained
by integrating the $c \bar c$ cross section from the $ c \bar c$
threshold, $\tau_1 = 2m_c/\sqrt s$, to the open-charm threshold,
$\tau_2 = 2 m_D/\sqrt s$. The differential cross section is then given as

\begin{eqnarray}
\frac{d\sigma}{dx_F}=F\int_{\tau_1}^{\tau_2} 2 \tau
d\tau H_{BT}(x_1,x_2;m^2)/(x_F^2+4\tau^2)^{1/2},
\label{eq:eq5}
\end{eqnarray}
where the factor $F$ signifies the fraction of the $c \bar c$ pairs
produced below the open-charm threshold to emerge as $J/\psi$.

Despite its simplicity, the
color-evaporation model is capable of describing many salient features of
hadronic $J/\psi$ production, including the shape of $d\sigma / dx_F$
as well as their beam-energy and
beam-type dependencies~\cite{Barger,Falciano,Vogt}. These
quantities are sensitive to hadron's parton distributions,
which govern the relative
contributions of $q \bar q$ annihilation and $g g$ fusion.
The success of the color-evaporation model suggests that
this model provides a useful tool for studying parton
distributions in pion and kaon from the $J/\psi$ production 
data~\cite{RefA,RefA1,RefB,RefC,RefD,RefE}.

The $gg$ fusion cross sections
for $J/\psi$ production in $\pi^- + p$ and $\pi^+ + p$ are identical.
This is a consequence of the charge symmetry at the partonic level~\cite{Tim}, 
which requires identical gluon
distributions in $\pi^-$ and $\pi^+$. Therefore, the difference
between the $J/\psi$ cross sections in $\pi^- + p$ and $\pi^+ + p$ interactions
solely comes from the $q \bar q$ annihilation contribution, namely,

\begin{eqnarray}
\centering
\sigma_{J/\psi}^{q\bar q}(\pi^- + p) & \propto & V_\pi(x_1) [u_p(x_2) +
\bar d_p(x_2)]\nonumber \\
& + & S_\pi(x_1)[2V_N(x_2)+4S_N(x_2)]; \nonumber \\
\sigma_{J/\psi}^{q\bar q}(\pi^+ + p) & \propto & V_\pi(x_1) [d_p(x_2) +
\bar u_p(x_2)]; \nonumber \\
& + & S_\pi(x_1)[2V_N(x_2)+4S_N(x_2)]. \nonumber
\end{eqnarray}
\begin{eqnarray}
S_\pi(x)  &=&  d_{\pi^+}(x) = \bar u_{\pi^+}(x)=u_{\pi^-}(x)=\bar d_{\pi^-}(x);
\nonumber \\
S_N(x)  &=&  [\bar u_p(x) + \bar d_p(x)]/2,
\label{eq:eq7}
\end{eqnarray}
where $S_\pi$ and $S_N$ are the sea quark distributions of the pion
and the nucleon, respectively. From Eq. (6) one can deduce that the
cross section difference between $\pi^- + p$ and $\pi^+ + p$ is
proportional to the product of pion's $V_\pi$ and proton's $u^V - d^V$
valence quark distributions:

\begin{equation}
\sigma_{J/\psi}(\pi^- + p)  -  \sigma_{J/\psi}(\pi^+ + p)) \propto  V_\pi(x_1) [u^V_p(x_2)-d^V_p(x_2)].
\label{eq:eq8}
\end{equation}
Since $V_\pi (x)$ is positive definite and $u^V_p(x)$ is larger than
$d^V_p(x)$ for $x>0.001$ (as seen in recent proton PDF global fit
parametrizations~\cite{CTEQ18,MHST20,NNPDF21}), Eq.~(7) shows that
production cross sections $\sigma_{J/\psi} (\pi^- + p)$ should be
greater than $\sigma_{J/\psi} (\pi^+ + p)$.

The NA3 Collaboration reported measurements of $\pi^- + p \to J/\psi +
x$ and $\pi^+ + p \to J/\psi + x$ using a 200 GeV pion
beam~\cite{NA3,Charpentier}.  Unseparated secondary beams were used,
and particle identification was performed using two Cherenkov counters
placed in the beam.  In the NA3 publication~\cite{NA3}, the $J/\psi$
cross sections were not listed directly. Nevertheless, the $\pi^- + p$
and $\pi^+ + p$ cross sections could be extracted from Figs 6a, 6c,
and 7b of Ref.~\cite{NA3}. An additional cross-check based on Fig. 36
of the thesis of Charpentier~\cite{Charpentier} gave consistent
results for the $\pi^- + p$ and $\pi^+ + p$ cross sections, shown in
Fig. 1(a).

\begin{figure}[tb]
\includegraphics*[width=0.5\textwidth]{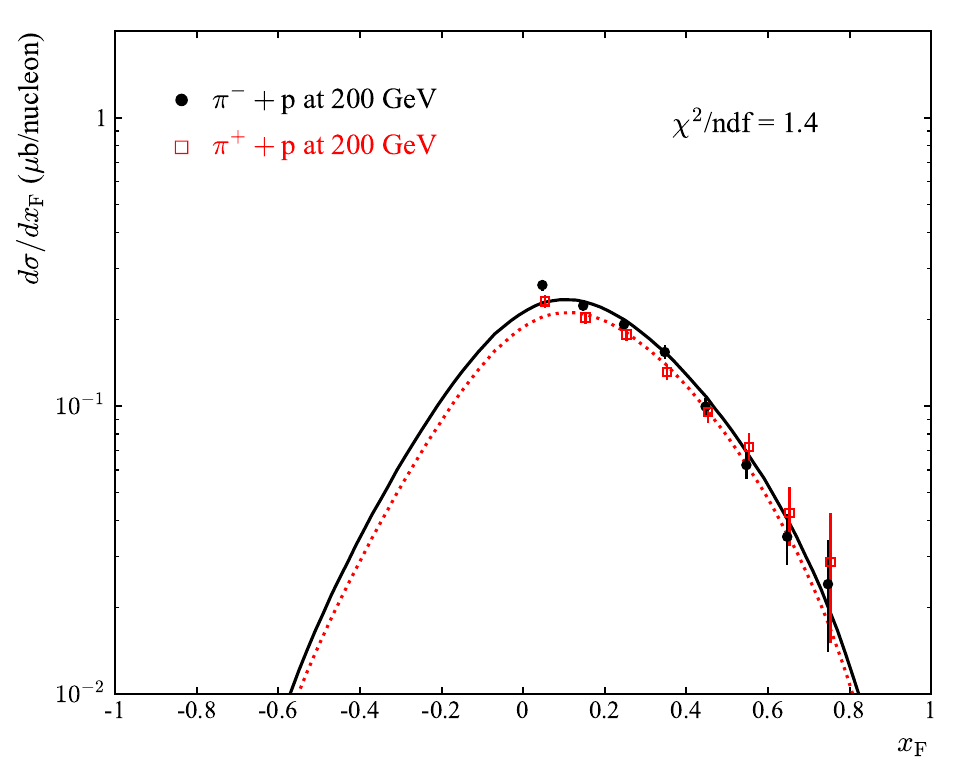}
\includegraphics*[width=0.5\textwidth]{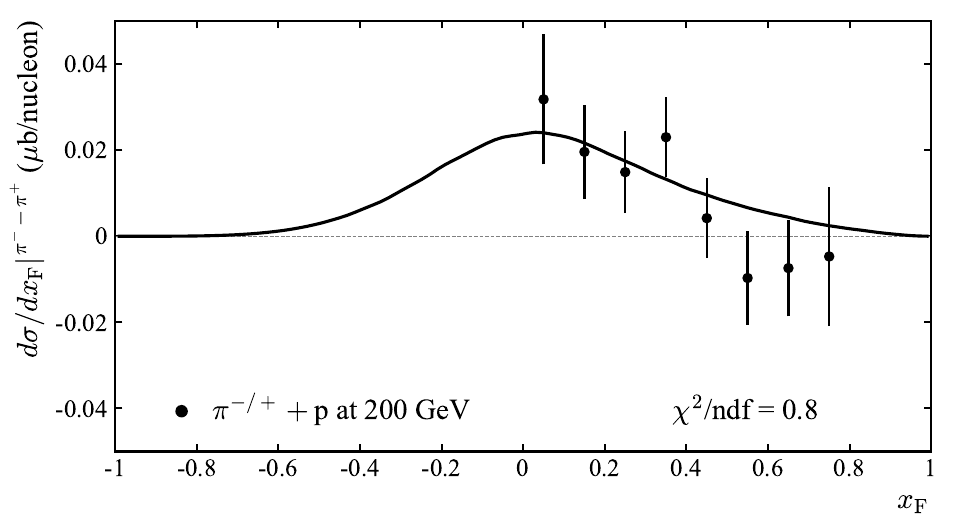}
\caption{(a): Measured cross sections for $\pi$-induced
$J/\psi$ production on proton at 200 GeV compared with calculations.
The data are from NA3~\cite{NA3,Charpentier} and the calculation
uses the NLO CEM model described in the text. The $\chi^2/\rm{ndf}$ refers to the simultaneous CEM fit to both $\pi^+$ and $\pi^-$ cross sections.
(b): The pion beam-charge asymmetry for $J/\psi$ production cross 
	sections compared with the NLO CEM calculation. 
}
\label{fig1}
\end{figure}

\begin{figure}[tb]
\includegraphics*[width=0.5\textwidth]{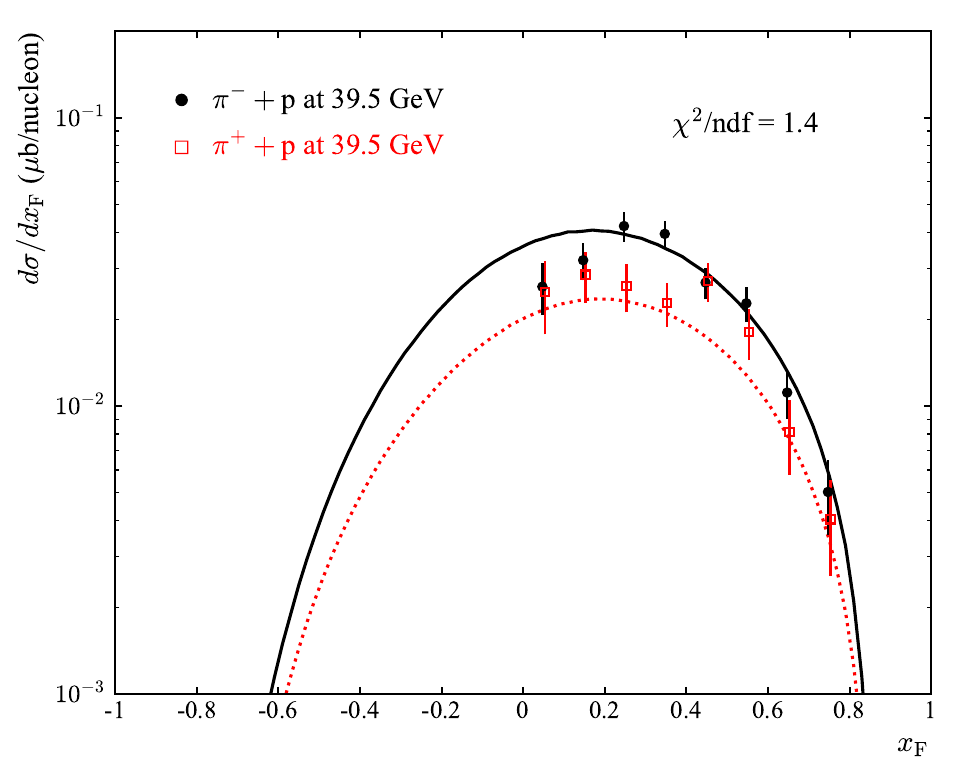}
\includegraphics*[width=0.5\textwidth]{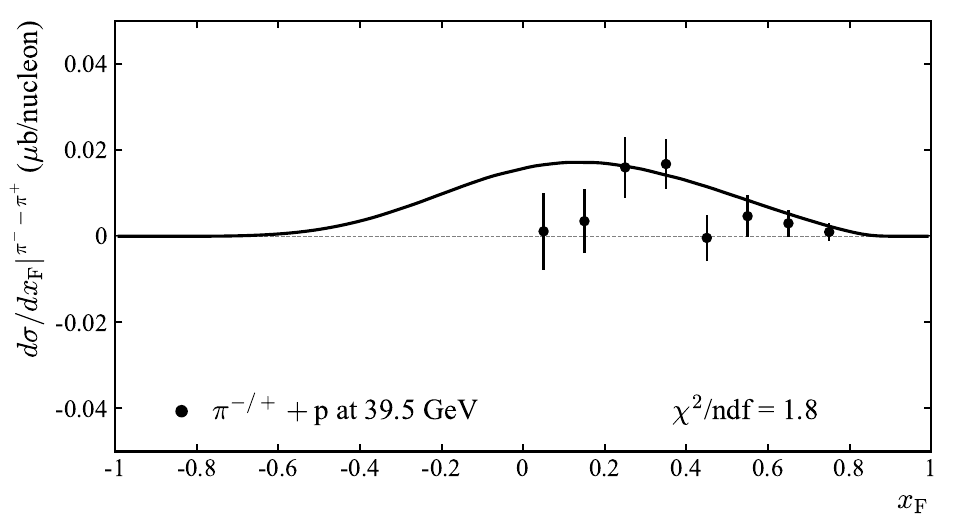}
\caption{(a): Measured cross sections for $\pi$-induced
$J/\psi$ production on proton at 39.5 GeV compared with calculations.
The data are from the WA39 experiment~\cite{WA39J} and the calculation
uses the NLO CEM model described in the text.
(b): The pion beam-charge asymmetry for $J/\psi$ production cross
	sections compared with the NLO CEM calculation.}  
\label{fig2}
\end{figure}

Figure 1(a) shows that $\sigma_{J/\psi} (\pi^- + p)$ is in general
greater than $\sigma_{J/\psi} (\pi^+ + p)$ in the measured kinematic
range of $0.0 < x_F < 0.8$, in agreement with the expectation based on
Eq. (7) discussed above. We have also performed the next-to-leading
order (NLO) calculations using the CEM
model~\cite{RefA,Nason,Mangano}. The solid and dotted curves in
Fig. 1(a) correspond to the calculations of $\pi^- +p$ and $\pi^+ +p$
$J/\psi$ production cross sections, respectively, using the
SMRS~\cite{SMRS} parton distribution (set 1) for the pion and the
CTEQ10NLO~\cite{CTEQ10NLO} parton distribution for the proton. The
hadronization factor $F$~\cite{RefA}, which represents the probability
of a $c \bar{c}$ pair fragmenting into charmonium in the CEM
calculation, is found to be 0.046. This value is obtained from a
simultaneous fit to the $x_F$ distributions of the cross sections for
both $\pi^-$ and $\pi^+$ induced reactions. The $\chi^2/\rm{ndf}$ of
the fit is 1.4. The data on the $\pi^+ + p$ and $\pi^- + p$ $J/\psi$
production cross sections are well described by the NLO CEM. We have
checked that using other proton PDFs, MSTW2008~\cite{MSTW2008} and
NNPDF23~\cite{NNPDF23} results in negligible differences. This
observation supports the assumption that in the $x$ region considered
the valence quark proton PDFs are well known.

\begin{figure}[tb]
\includegraphics*[width=0.5\textwidth]{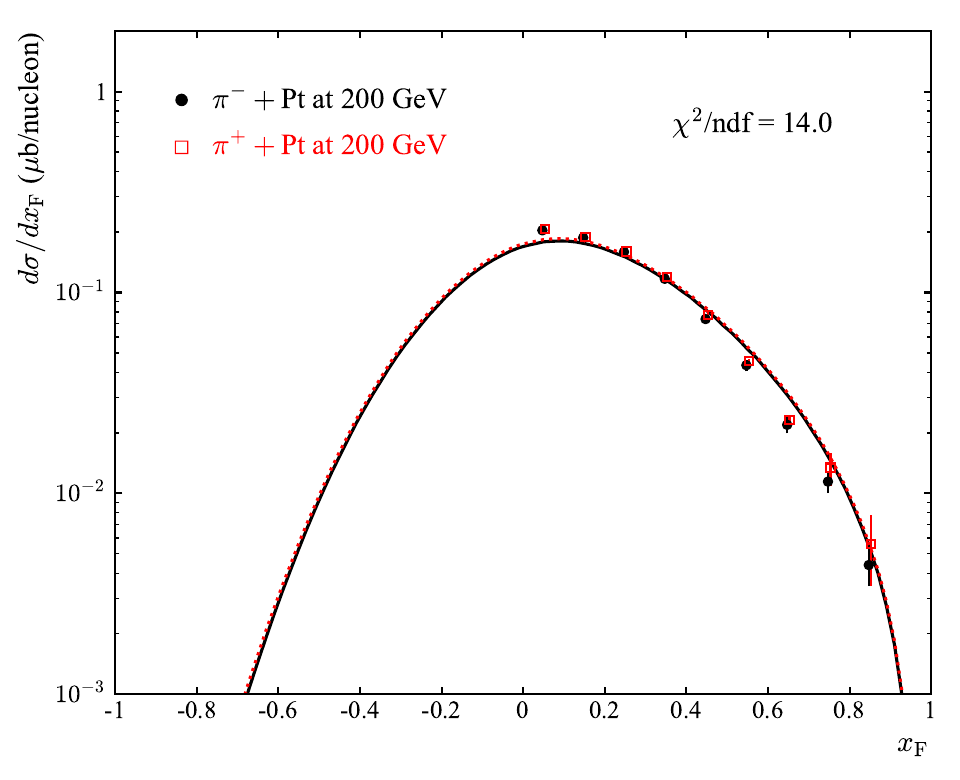}
\includegraphics*[width=0.5\textwidth]{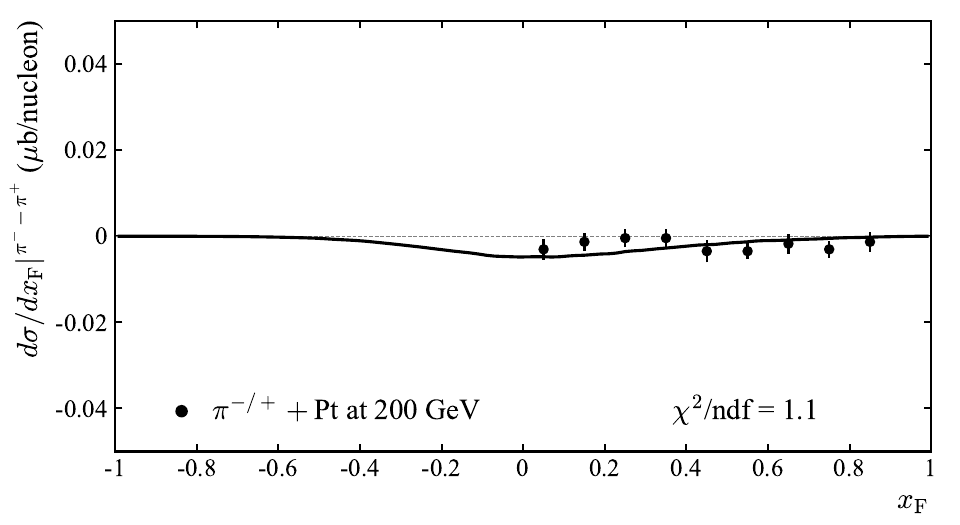}
\caption{(a): Measured cross sections per target nucleon for $\pi$-induced
$J/\psi$ production on a platinum target at 200 GeV compared with calculations.
The data are from the NA3 Collaboration~\cite{NA3,Charpentier} and 
the calculation
uses the NLO CEM model described in the text.
(b): The pion beam-charge asymmetry for $J/\psi$ production cross
sections compared with the NLO CEM calculation.} 
\label{fig3}
\end{figure}

Figure 1(b) displays $\sigma_{J/\psi} (\pi^- + p) - \sigma_{J/\psi}
(\pi^+ + p)$ for both the data and the NLO CEM calculations.
The qualitative agreement between the data and the calculation suggests that 
future high-statistics data on $J/\psi$ production of $\pi^\pm + p$
over a broad range of kinematics variables would be of
great interest. 
Of particular interest is the region of large $x_F$.
The current information for $V_\pi(x)$ at large $x$ is based largely on
the E615 $\pi^- + W$ Drell-Yan data~\cite{E615}. 
The presence of various nuclear 
effects at large $x$, such as the partonic energy loss effect
observed in proton-induced Drell-Yan experiments~\cite{Garvey,Arleo},
could introduce significant uncertainties in the extraction of  
$V_\pi(x)$ from these data. As nuclear effects are absent in $\pi + p$
$J/\psi$ production, these data would provide a valuable
and independent measurement of $V_\pi (x)$.
\begin{figure}[tb]
\includegraphics*[width=0.5\textwidth]{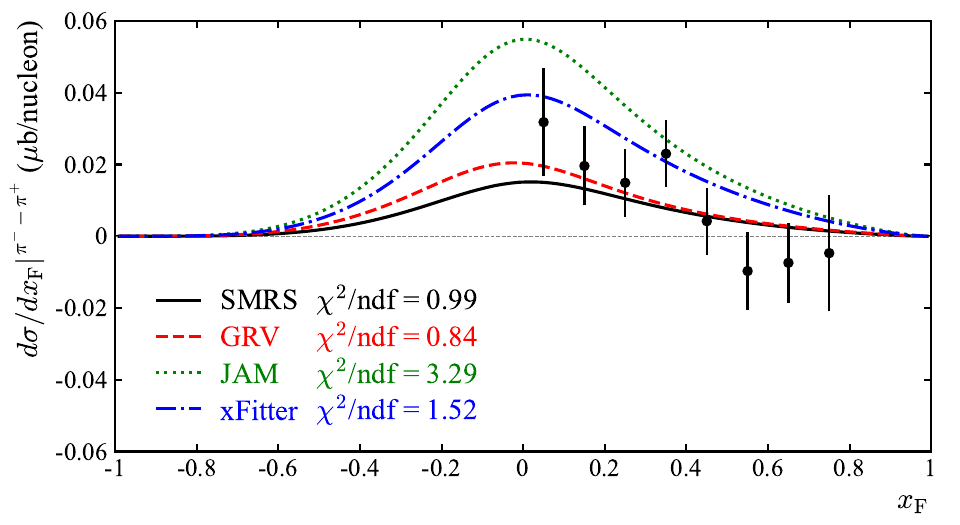}
\includegraphics*[width=0.5\textwidth]{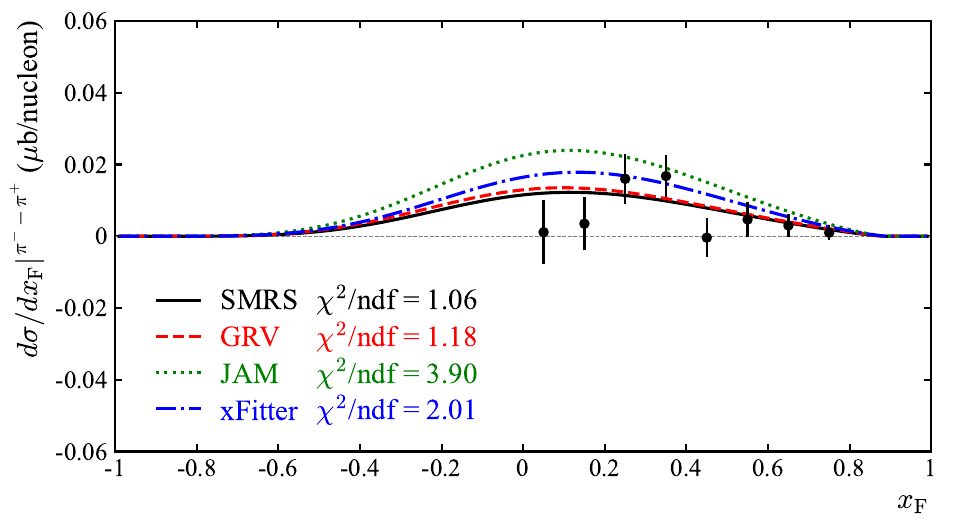}
\caption{(a): The beam-charge asymmetry for pion-induced $J/\psi$
production on proton at 200 GeV compared with NRQCD calculation
using four different pion PDFs.
(b): Same as (a), but for the WA39 data with 39.5 GeV
pion beam.}
\label{fig4}
\end{figure}

Other than the NA3 data discussed above, the only measurement of the
$\pi + p$ $J/\psi$ production cross section was performed by the WA39
Collaboration at a lower beam energy of 39.5 GeV~\cite{WA39J}.  The
data from the WA39 measurement are shown in Fig. 2 (a). It is
interesting to note that the data again agree with the expectation
that $\sigma_{J/\psi} (\pi^- + p) > \sigma_{J/\psi} (\pi^+ + p)$ for
the entire measured kinematic range of $0.0 < x_F < 0.7$. The solid
and dotted curves in Fig. 2(a) show the NLO CEM calculations of $\pi^-
+p$ and $\pi^+ +p$ cross sections using the same pion and proton PDFs
as in Fig. 1. Again, the hadronization factor, $F=0.071$, is
determined by a simultaneous fit of both the $\pi^- + p$ and $\pi^+ +
p$ data and the $\chi^2/\rm{ndf}$ of the fit is 1.4. The value of $F$
is larger than that for the data at 200 GeV. Such energy dependence of
the hadronization factor is consistent with our previous
study~\cite{RefA} performed with the CEM model. Figure 2(b) shows the
qualitative agreement between the data and the NLO CEM calculations
for the $\pi^-$ and $\pi^+$ cross section difference.
At the relatively low beam energy of 39.5 GeV, the $gg$ fusion subprocess is less important than at 200
GeV~\cite{RefA}. Nevertheless, Eq. (7) shows that the cross section difference is independent of the gluon distributions and only depends on the valence
quark distributions of the pion and the proton. This explains the similarity
between Fig. 1(b) and Fig. 2(b), suggesting a weak beam-energy dependence
of the $\pi^+$ and $\pi^-$ cross section difference.

While Eq. (1) shows that the Drell-Yan cross section difference between
$\pi^- + D$ and $\pi^+ + D$ is proportional to $V_\pi (x_1) V_N(x_2)$, it
is interesting to note that the corresponding cross section difference
for $J/\psi$ production is expected to vanish, namely,
\begin{eqnarray}
\sigma_{J/\psi} (\pi^- + D) - \sigma_{J/\psi} (\pi^+ + D)= 0.
\label{eq:eq9}
\end{eqnarray}
This result can be readily obtained from Eq. (6), together with 
analogous expressions for $\pi^- + n$ and $\pi^+ + n$ cross sections 
assuming isospin symmetry. The notable difference between Eq. (1) and
Eq. (8) reflects the distinct natures of the Drell-Yan process and
the $J/\psi$ production. While the $q\bar q$ annihilation for $J/\psi$
production (Eq. (6)) proceeds via strong interaction, the $q\bar q$ 
annihilation for the Drell-Yan process is an electromagnetic
process. The absence of the $Q^2_i$ dependence, where $Q_i$ is the
charge of quark $i$, in the $J/\psi$ production cross section 
accounts for the difference between Eq. (1) and Eq. (8).

The NA3 Collaboration has also measured $J/\psi$
production with $\pi^+$ and $\pi^-$ beams on a platinum 
target~\cite{NA3,Charpentier}.
For a neutron-rich nuclear target like platinum, the beam-charge
asymmetry for pion-induced $J/\psi$ production is expected to be
opposite to that for the proton target. Indeed, Eq. (7) can be
generalized into the case for a target nucleus consisting of
$Z$ protons and $N$ neutrons as follows:
\begin{eqnarray}
\sigma_{J/\psi}(\pi^- + \ce{^{N}_{Z}A})
- \sigma_{J/\psi}(\pi^+ + \ce{^{N}_{Z}A})~~~~~~~~~~\nonumber \\
\propto \frac{Z-N}{A} V_\pi(x_1) [u^V_p(x_2)-d^V_p(x_2)],
\label{eq:eq10}
\end{eqnarray}
where $A=Z+N$ and $\sigma$ refers to the per-nucleon cross section.
For the neutron-rich platinum nucleus, $Z<N$, and Eq. (9) implies a
negative beam-charge asymmetry. As shown in Fig. 3, the NA3 $\pi$ + Pt
$J/\psi$ production data indeed has a negative asymmetry consistent
with this expectation. Eq. (9) shows that the magnitude of the
asymmetry for the platinum target should be suppressed by a factor of
$(Z-N)/A = -0.2$ relative to that for the proton target. This
expectation is confirmed from a comparison between Fig. 3 and Figs. 1
and 2, and is in agreement with the theoretical calculation using the
HKNnlo nuclear PDF from Ref.~\cite{HKNnlo}.


To further illustrate the sensitivity of the beam-charge asymmetry of
the $J/\psi$ production data to pion PDFs, we show in Fig. 4 the
comparison between the data with calculations in the framework of
NRQCD (Non-Relativistic QCD), using four different pion
PDFs~\cite{SMRS,GRV,JAM,xFitter}. The Long Distance Matrix Elements
(LDMEs) in the NRQCD calculations were taken from an earlier analysis
with a global fit to existing $\pi^-$-induced and proton-induced
$J/\psi$ production data~\cite{RefC}. The $\pi^+$-induced $J/\psi$
production data were not included in this global fit. Figure 4 shows
that the beam-charge asymmetry data at both beam energies are better
described by the SMRS~\cite{SMRS} and GRV~\cite{GRV} pion PDFs than
the JAM~\cite{JAM} and xFitter~\cite{xFitter} PDFs. To examine the
sensitivity of the results shown in Fig. 4 to the proton PDFs, we have
repeated the calculations by using two other sets of proton PDFs
(MSTW2008 and NNPDF23) and the results are practically independent of
the proton PDFs used in the calculation. The insensitivity to the
proton PDFs supports our finding that the beam-charge asymmetry in
pion-induced $J/\psi$ production is a viable experimental tool to
access the pion PDFs.

We note that measurements of the beam-charge asymmetry in 
pion-induced $J/\psi$ production on a hydrogen target can also probe the 
proton PDFs. In particular, 
Eq. (8) implies that, once the values of $V_\pi(x_1)$ are well
determined, the $\pi + p$ $J/\psi$ production data can provide new
information on proton's $u_p^V(x) - d_p^V(x)$ valence quark distribution.
Current knowledge on $u_p^V(x) - d_p^V(x)$ is rather poor at large value
of $x$. The $\pi + p$ data are free
from the uncertainty of nuclear effects encountered in the tagged-DIS 
measurements on a deuterium target~\cite{CLAS12,CLAS14} 
or the DIS experiment on $A=3$ targets~\cite{Marathon}.

In summary, we have shown that the beam-charge asymmetry of pion-induced
$J/\psi$ production on a hydrogen target, 
$\sigma_{J/\psi} (\pi^- + p) - \sigma_{J/\psi} (\pi^+ + p)$, has
a positive sign and is sensitive to the product of the pion's and the 
proton's valence quark distributions. We have compared existing data on the
beam-charge asymmetry for several different target nuclei with the theoretical
calculations. The good
agreement between data and calculation suggests that the beam-charge 
asymmetry for pion-induced $J/\psi$ production is a viable
method to study the valence quark distributions of pion and proton.
The much larger cross section for $J/\psi$ production compared with the 
Drell-Yan represents a significant advantage. New data covering a broad 
kinematic region with high statistics, as anticipated in future experiments 
at AMBER~\cite{Amber}, would be of great interest.

This work is supported in part by the U.S. National Science Foundation
and the Ministry of Science and Technology of Taiwan.

\end{document}